\documentstyle[12pt]{article}
\begin{document}
\mbox{~}
\vspace{0.1cm}

\begin{center}
{\Large\bf Relational interpretation of the wave function\\[0.2cm]
 and a possible way around Bell's theorem}\\[1.9cm]
{\large\bf Thomas Filk }\\[0.6cm]
{\large Institute for Theoretical Physics,
Universit\"at Freiburg, Hermann-Herder-Str.\ 3,
D-79094 Freiburg, Germany \\[0.3cm]
Parmenides Foundation; Via Mellini 26-28, 
57031 Capoliveri, Italy \\[0.3cm]
Institute for Frontier Areas in Psychology and Mental Health; 
Wilhelmstr.\ 1a; 79098 Freiburg, Germany} 
\end{center}
\vspace{1.2cm}

\thispagestyle{empty}
\setcounter{page}{0}
\begin{abstract}
The famous ``spooky action at a distance''
in the EPR-szenario is shown to be a local interaction, once
entanglement is interpreted as a kind of ``nearest neighbor''
relation among quantum systems. Furthermore, the
wave function itself is interpreted as encoding the
``nearest neighbor''
relations between a quantum system and spatial points. 
This interpretation becomes natural, if we view space
and distance in terms of relations among spatial
points. Therefore, ``position'' becomes a purely relational
concept. This relational picture leads to a new perspective
onto the quantum mechanical formalism, where many of the
``weird'' aspects, like the particle-wave duality, the
non-locality of entanglement, or the ``mystery'' of
the double-slit experiment, disappear. Furthermore, this
picture cirumvents the restrictions set by Bell's inequalities, 
i.e., a possible (realistic) hidden variable theory
based on these concepts can be local and at the same
time reproduce the results of quantum mechanics.
\vspace{0.3cm}

\end{abstract}

\newpage

\section{Introduction}

For many people, our quantum world still 
encompasses a few ``mysteries'':
\begin{itemize}
\item[-]
How can an object behave like a pointlike 
particle in some cases and like an extended wave in
other cases? (Particle-wave duality)
\item[-]
In particular,
how can an object (say, an elementary particle) appear
pointlike whenever it is measured directly, but on the
other hand appear to be at two (or several) places at
the same time when it is not observed (like the electron
in the double slit experiment or the photon in
a Mach-Zehnder interferometer)?
\item[-]
How can the results of measurements on two entangled particles
be correlated even when they are ``miles away'', although,
according to the formalism of quantum mechanics, 
we are not allowed to assume that the output of these
measurements is predetermined in adavance 
(Einstein's spooky action at a distance).
\end{itemize}
Furthermore, for all those, who favour an 
``objective and realistic'' interpretation of the fundamental
principles of our world, Bell's inequalities imply a
major draw-back. The experimental evidence in favor of
quantum mechanics and the violation of Bell's inequalities
in our physical world is overwhelming and no longer a 
matter of serious debates. The generally accepted 
conclusion is that the non-deterministic aspects of
quantum mechanics are fundamental, i.e., already the
assumption of a ``hidden variable'' determining the output
of certain experiments in advance leads to contradictions
unless we give up locality. Hence, all reformulations of
quantum mechanics based on realism, like e.g.\
Bohm's quantum mechanics, include non-local interactions.
The apparent discrepancy with the theory of relativity can only 
be overcome by proving that these non-local (hidden) interactions
cannot be used for information transfer and should,
therefore, not be interpreted as signals. However, the
``spooky action at a distance'', as Einstein called it,
remains.

At a closer inspection, all the problems mentioned above
are in some way or another related to what we mean by
``locality'', ``position'', ``distance'', ``place'', and
in particular ``to be somewhere''.
All definitions of locality (including the 
precise definition provided by the algebraic formalism of 
quantum mechanics) are based on a classical concept of space-time 
and may, therefore, not be completely consistent. ``Position'' refers
to the points of a fixed background space. On the other
hand, we expect that the concept of such a background 
space emerges as the classical limit of an underlying
quantum theory of space and time. So, our present formulation
of quantum theory
is of a somewhat hybrid nature in that it describes 
quantum objects as being ``embedded in'' or 
``living on'' a classical space. Although the
interactions among the quantum objects are treated in
a quantum mechanical way, the concepts of space and space-time and,
in particular, the relations between quantum objects
on the one side and space-time on the other side 
are footed on a classical description. 

One might object by pointing out that Planck's scale is about
25 orders of magnitude away from atomic scales and, therefore,
should not play any role in the quantum world as we see
it today. However, there is at least one other 
structure of space (and space-time) which survives the
25 orders of magnitude: the metric field in special and 
general relativity. Space-time is more than a simple set of
events, otherwise it would be impossible to measure and
compare distances at different space-time points.
This structure is usually taken for granted, but 
it should be viewed as the large scale remnant of some 
unknown underlying structure of quantum space-time. 

In this article, I will argue that already a simple reformulation
of spatial concepts in terms of a relational interpretation
might give us a new understanding of ``locality''. This
concept does not only overcome the seeming discrepancy between 
realism and the violation of Bell's inequalities in 
our quantum world, but it almost trivially explains the other
weird aspects mentioned above, like the particle-wave
duality, the double-slit experiment etc. Although 
these concepts should rather be formulated 
in terms of space-time events and not in terms of spatial points,
the major part of this article refers to a ``relational theory of
space'', and only at the end there will be a few remarks about
a generalization towards a ``relational theory of space-time''. 

The expression ``relational quantum mechanics'' already exists
in the literature and refers to Rovelli's (and others) 
interpretation of quantum mechanics \cite{Rovelli},
according to which states or the results of measurements do 
not have an absolute meaning, but all these 
concepts of quantum mechanics are to be understood
in relation to the state of an observing system. (This
interpretation is close to Everett's ``relative state''
interpretation of quantum mechanics \cite{Everett}, although
the two interpretations differ with respect to the
``many world'' aspect which deWitt later attached to
Everett's interpretation \cite{DeWitt}.)

In a way, the present approach can be considered as
a refinement or extension of Rovelli's concept. In
relational quantum mechanics the state of a system is
not attributed to the system itself (like in ontic
interpretations) or to the observing system 
(like in epistemic interpretations), but the state
is rather attributed to the boundary (or ``cut'') between
the quantum system and the rest of the world, and it contains
the information or knowledge which the rest of the world in principle
has about the quantum system due to past interactions
and present correlations between these two systems.
The ``refinement'' discussed in the present article 
consists in the observation that in a world with
relational space structure, the ``cut'' could be placed 
between the particle and space, i.e., the wave function
$\psi(x)$ describes the ``information'' which
space has about the particle. In general relativity, 
space-time has its own degrees of freedom which interact with 
other objects from which it can be deprived, so why not
deprive the objects from space-time and consider
them as entities of their own? ``Position'' should not 
be treated as an embedding of a particle, but 
as an external property arrising from the relations
between this particle and those objects which make up
space-time.       

In the next section \ref{sec2}, I give a brief introduction to 
models of relational spaces and to the general ideas
of how a relational picture for quantum objects and spatial
points may influence our understanding of quantum mechanics.
Even if some of the more specific ideas elaborated in
the following sections should turn out to be wrong, the
general scenario explained in section \label{sec2} may 
still be true. In section \ref{sec3}, I will specify
the ideas by making some assumptions about the type
of relations between quantum objects and spatial points.
Section \ref{sec4} describes a simple model for the propagation
of relations and introduces a new concept of locality.
The essential purpose of sections \ref{sec3} and \ref{sec4}
is to show that the relational picture fits well into the
present formulation of quantum mechanics and that there
is no need to change the formulas but to change the
interpretations.
In section \ref{sec5}, I will raise the question whether
all amplitudes in quantum mechanics express relations,
and I will give some examples of relational concepts
in todays standard model of elementary particles. 
Section \ref{sec7} contains
a few remarks about the extension of ``particles in a 
relational space'' to a model of ``events in relational 
space-time''. A brief summary concludes this article.
 
\section{Relational space}
\label{sec2}

The clearest and most uncompromising formulation of a
relational theory of space has been put forward by 
Ren\'e Descartes in his ``Principles of Philosophy'', 
published in 1644 \cite{Descartes}. In the second part
of his ``Philosophiae'', entitled ``About the Principles
of Corporal Things'', Descartes
argues that there is no such thing as empty space, and
our whole concept of space is just an abstraction
of what in reality are ``relations'' between bodies. For
Descartes these relations express ``immediate neighborship''  
or the ``contact'' of bodies. ``Movement'' is merely a change 
of these relations, i.e.\ a rearrangement of objects. The 
``bodies'' may not always be visible to us, but ``if God 
would remove from a vessel all the body''  
then ``the sides of the vessel would thus come into proximity
with each other''.

Later, similar concepts have been put forward by Gottfried
Wilhelm Leibniz. In his famous exchange of letters with 
Samuel Clarke \cite{Clarke}, he puts his concepts of a relational
space and relational time against Newton's concepts of
absolute space and absolute time. For Leibniz, space is 
just an abstraction of ``the order of coexistences''. 
Leibniz is less
clear about the nature of the relations between physical
objects and refers to them as ``some relation of distance'' 
(in a different context he speaks about ``perceptions'' 
\cite{Monadologie}), but the fundamental ideas are similar to 
those of Descartes. Towards the end of the 19th century it was 
Ernst Mach who brought up the subject again \cite{Mechanik}. 
His work has greatly influenced 
Einstein in his development of the general theory of relativity. 

More recently, the importance of a relational formulation
of the fundamental laws of physics has been emphasized by 
Julian Barbour (amongst others), who, together with B.~Bertotti, 
constructed a theory of mechanics based on  
relational principles \cite{Barbour}. In other approaches,
the microscopic structure of space-time is modelled 
in terms of relational principles, like the causal sets
of R.~Sorkin \cite{Sorkin}.

How does a relational space look like? Imagine 
certain objects which, for simplicity, will be represented 
by points, keeping in mind, however, that any notion of
``extension in space'' for an elementary object is 
meaningless in the relational view. 
Also for simplicity, I will distinguish objects 
representing ``spatial points'' and objects
representing particles. In this section,  
I will consider only one type of relation.
This relation is represented by a line. This leads to 
the concept of a graph. The relation is either present 
or not present. Mathematically, this can be expressed by
the adjacency matrix:
\begin{equation}
\label{eq1}
       A(x,y) = \left\{ \begin{array}{ll}
       1 \mbox{ if $x$ and $y$ are related}\\
       0 \mbox{ otherwise} \end{array} \right. ~.
\end{equation}
In the next section, I will add some hypotheses about 
the nature of the relations, and the
concept of an adjacency matrix will be generalized.

The spatial points are related in such a way that,
on large scales, space looks like a three dimensional
manifold. A three dimensional lattice has this property,
however, there is no reason (and no necessity) to assume that 
space at the fundamental level resembles a regular lattice.

The distance between two points is defined purely
intrinsically as a suitable average of the lengths of
paths connecting the points. The length of a path
is given by the number of lines of this path. It will
be important to notice that the (macroscopic) distance
between points is not determined by the length of the 
shortest path alone, but by a suitable average over all paths
(where a possible weighting of paths might depend on 
the length of a path). I will say more about ``distance''
later.

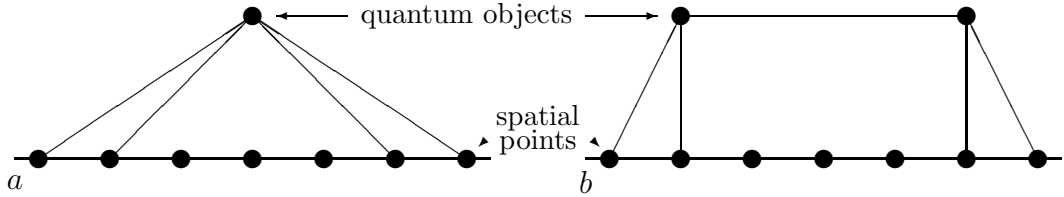
\begin{figure}
\unitlength0.9pt
\begin{picture}(450,100)(0,0)
\put(10,10){\circle*{8}}
\put(40,10){\circle*{8}}
\put(70,10){\circle*{8}}
\put(100,10){\circle*{8}}
\put(130,10){\circle*{8}}
\put(160,10){\circle*{8}}
\put(190,10){\circle*{8}}
\put(100,70){\circle*{8}}
\put(10,10){\line(3,2){90}}
\put(190,10){\line(-3,2){90}}
\put(40,10){\line(1,1){60}}
\put(160,10){\line(-1,1){60}}
\put(0,10){\line(1,0){200}}
\put(190,70){\makebox(0,0){\small quantum objects}}
\put(220,27){\makebox(0,0){\small spatial}}
\put(220,17){\makebox(0,0){\small points}}
\put(140,70){\vector(-1,0){30}}
\put(240,70){\vector(1,0){30}}
\put(205,25){\vector(-1,-1){10}}
\put(236,25){\vector(1,-1){10}}
\put(250,10){\circle*{8}}
\put(280,10){\circle*{8}}
\put(310,10){\circle*{8}}
\put(340,10){\circle*{8}}
\put(370,10){\circle*{8}}
\put(400,10){\circle*{8}}
\put(430,10){\circle*{8}}
\put(280,70){\circle*{8}}
\put(400,70){\circle*{8}}
\put(250,10){\line(1,2){30}}
\put(430,10){\line(-1,2){30}}
\put(280,10){\line(0,1){60}}
\put(400,10){\line(0,1){60}}
\put(240,10){\line(1,0){200}}
\put(280,70){\line(1,0){120}}
\put(0,0){\makebox(0,0){$a$}}
\put(240,0){\makebox(0,0){$b$}}
\end{picture}
\label{fig1_1}
\caption{(a) A relational particle can be at two
places at the same time. (b) Two particles which are
far apart with respect to the spatial distance 
can still be immediate neighbors.}
\end{figure}

The ``position'' of an object is defined by its relation
to spatial points. Therefore, the position of an 
object can be localized or extended, depending
on whether this object has only relations to points which
are close to each other or far apart from each other with
respect to the intrinsic distance. It is
even conceivable that a single object can be at two 
``places'' simulateously (see fig.~\ref{fig1_1}\,a), so the
relational picture overcomes ``the only mystery'' (this
expression is due to Feynman \cite{Feynman}) of quantum 
mechanics. In this picture, a particle can 
be related to two places at the same time which may
allow for a natural explanation of
the double slit experiment. Notice that
only the relations matter, not the ``length'' of a line
or the ``position'' of an object in a graphical representation.

Next, imagine two particle-like objects each having
its relations to spatial points. With
respect to the intrinsic relations among spatial points, these
two particles could be far apart from each other. However,
these particles may be immediate neighbors with respect
to other relations (in
the next section, I will argue that measures of entanglement are 
a good candidate for the relations between particles, but
for the moment I will not make any assumptions about the nature of
these relations).
With respect to this relation the particles are
``immediate neighbors'' (see fig.~\ref{fig1_1}\,b). 
Performing an experiment on
one of the particles can immediately influence the
other without violating locality, even though the
particles look ``miles apart'' from each other with
respect to their spatial relations. (In sec.\ \ref{sec4}
the concept of locality is generalized to the dynamics
of relations.)

In order that two objects are observed to have a large
spatial distance even
thought they are directly related to each other 
(and therefore there exists a path of length 1 connecting 
these objects), it is
important that not only the shortest path on the graph
contributes to the distance, but all paths. 
The fact that there exists a short-cut
(via the relation between the particle-like objects) does not
significantly change the macroscopic distance between 
two spatial points, which is obtained by a suitable
averaging procedure. A huge number of such short-cuts, 
stemming, e.g., from many particles, could alter the 
macroscopically observable spatial distance between 
spatial points. (This observation might even be a starting
point for bringing in general relativity. That the presence
of matter changes the metric is one more feature which
looks more natural in a relational theory of space. 
However, further conclusions are still very speculative.)
  
Thus, we have seen that the relational picture solves
two of the major mysteries of quantum mechanics almost
trivially: the fact that a particle can ``be in'' (in the
sense of ``have relations to'') more than
one place at the same time, and the ``spooky action
at a distance'' in the EPR-scenario. 

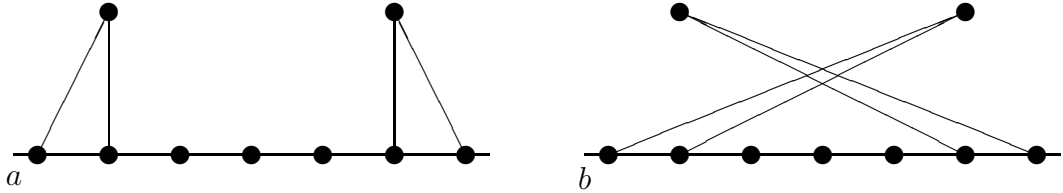
\begin{figure}
\unitlength0.9pt
\begin{picture}(450,100)(0,0)
\put(10,10){\circle*{8}}
\put(40,10){\circle*{8}}
\put(70,10){\circle*{8}}
\put(100,10){\circle*{8}}
\put(130,10){\circle*{8}}
\put(160,10){\circle*{8}}
\put(190,10){\circle*{8}}
\put(40,70){\circle*{8}}
\put(160,70){\circle*{8}}
\put(10,10){\line(1,2){30}}
\put(190,10){\line(-1,2){30}}
\put(40,10){\line(0,1){60}}
\put(160,10){\line(0,1){60}}
\put(0,10){\line(1,0){200}}
\put(250,10){\circle*{8}}
\put(280,10){\circle*{8}}
\put(310,10){\circle*{8}}
\put(340,10){\circle*{8}}
\put(370,10){\circle*{8}}
\put(400,10){\circle*{8}}
\put(430,10){\circle*{8}}
\put(280,70){\circle*{8}}
\put(400,70){\circle*{8}}
\put(250,10){\line(5,2){150}}
\put(430,10){\line(-5,2){150}}
\put(280,10){\line(2,1){120}}
\put(400,10){\line(-2,1){120}}
\put(240,10){\line(1,0){200}}
\put(0,0){\makebox(0,0){$a$}}
\put(240,0){\makebox(0,0){$b$}}
\end{picture}
\label{fig1_2}
\caption{(a) Two identical particles having relations
to different spatial points. (b) The exchange of the two
particles leads to the identical set of relations.}
\end{figure}

There is a third aspect of quantum mechanics which 
appears more natural in the relational picture.
When the spatial relations of two identical objects are 
exchanged (see fig.\ \ref{fig1_2}), the set of relations
remains unchanged, i.e., the two situations not
only look identical but they are identical. This is
how they are treated in quantum mechanics.
 
\section{The wavefunction as the ``adjacency matrix'' of
relations}
\label{sec3}

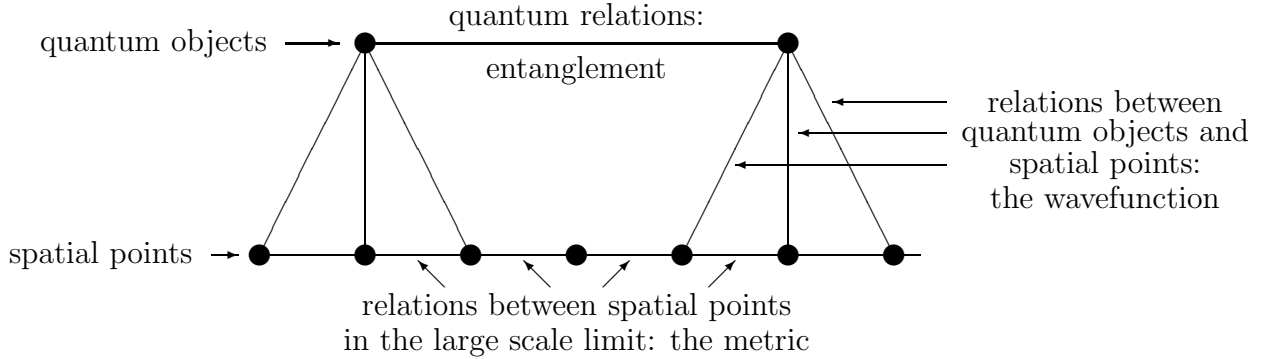
\begin{figure}
\begin{picture}(450,150)(-75,-30)
\put(10,20){\circle*{8}}
\put(50,20){\circle*{8}}
\put(90,20){\circle*{8}}
\put(130,20){\circle*{8}}
\put(170,20){\circle*{8}}
\put(210,20){\circle*{8}}
\put(250,20){\circle*{8}}
\put(50,100){\circle*{8}}
\put(210,100){\circle*{8}}
\put(10,20){\line(1,2){40}}
\put(250,20){\line(-1,2){40}}
\put(90,20){\line(-1,2){40}}
\put(170,20){\line(1,2){40}}
\put(50,20){\line(0,1){80}}
\put(210,20){\line(0,1){80}}
\put(10,20){\line(1,0){250}}
\put(50,100){\line(1,0){160}}
\put(130,110){\makebox(0,0){quantum relations:}}
\put(130,90){\makebox(0,0){entanglement}}
\put(130,0){\makebox(0,0){relations between spatial points}}
\put(130,-13){\makebox(0,0){in the large scale limit: the metric}}
\put(330,78){\makebox(0,0){relations between}}
\put(330,66){\makebox(0,0){quantum objects and}}
\put(330,54){\makebox(0,0){spatial points:}}
\put(330,42){\makebox(0,0){the wavefunction}}
\put(-30,100){\makebox(0,0){quantum objects}}
\put(-50,20){\makebox(0,0){spatial points}}
\put(20,100){\vector(1,0){20}}
\put(-8,20){\vector(1,0){10}}
\put(80,8){\vector(-1,1){10}}
\put(120,8){\vector(-1,1){10}}
\put(140,8){\vector(1,1){10}}
\put(180,8){\vector(1,1){10}}
\put(270,78){\vector(-1,0){43}}
\put(270,66){\vector(-1,0){56}}
\put(270,54){\vector(-1,0){80}}
\end{picture}
\label{fig2_1}
\caption{Three types of relations are present in our formulation of
quantum mechanics. Relations among quantum objects are due to
entanglement, relations among spatial points should in a large scale
limit lead to the metric, and relations between quantum objects and
spatial points are described by the wave function.}
\end{figure}

In this section, I will make the relational picture more
concrete by adding a few hypotheses about the nature of the
relations. This will lead to the simplified picture which
is sketched in fig.\ \ref{fig2_1}. 

According to this picture, our present formulation of 
quantum mechanics includes essentially three types of 
relations, although we are usually aware of only one.
The first (and well-known) relation emerges in the
description of the interaction between quantum systems 
like elementary particles. Such interactions, in the
context of quantum field theory
attributed to the exchange of other particles, lead
to entanglement which can be viewed as a quantum
relation among quantum systems. Hence, quantum
mechanics describes the relations among quantum objects.
This aspect of quantum mechanics is not 
put into question. 
What is new, however, is the idea to consider 
entanglement as a ``nearest neighbor'' relation.
Whenever two particles or two systems are 
entangled, they are, in a sense to be defined,
immediate neighbors.

One might object that entanglement is not
a binary relation. The information about an arbitrary 
entangled state of more than two particles cannot
be decomposed into pairwise relations and thus cannot
be represented by a simple graph. This is true and
might hint at the possibility that the relational picture
presented in this article has to be generalized from
simple graphs to, for instance, abstract simplicial 
structures, including also ternary and higher order 
relations. Despite very intensive research, a satisfying 
theory for measures of entanglement is still missing
(see e.g.\ \cite{Hill,Wootters,Horodecki,Bruss} and
references therein.) A
characterization by graphs has been attempted in 
\cite{Plesch}, and, more recently, Bob Coecke found a
formulation in which entanglement can be characterized
as a generalized relation in a categorial sense \cite{Coecke}. 
A more detailed theory of entanglement might 
influence the details of the framework presented here, 
but the general ideas remain untouched.   

The second type of relations is usually not 
mentioned explicitly, but it is inherent in 
the description. These are the relations among spatial 
points (or space-time points, i.e.\ events).
As we are not yet in the possession of a fundamental 
quantum theory of space and time, we do not know the 
nature of these relations on Planck's scale. Possible
candidates are Rovelli's and Smolin's ``loop space'' \cite{Smolin} or 
an extension of the spin networks of Penrose\cite{Penrose}. 
Whatever the theory at Planck's scale may look like, 
in a large scale limit these relations should 
lead to the emergence of distance and to the metric field 
$g_{\mu \nu}$ of general relativity. In lack of the 
fundamental theory, I will adopt here the simple
relational model of the previous section:   
Spatial points are represented by the vertices of a network (or graph)
and the relations between these spatial points will be
represented by the lines of the graph. 
This model is to be understood as a 
``semi-phenomenological''
description of space at small distances.

The third type of relations is usually not even mentioned
implicitly, but may be the most important one for our 
understanding of quantum mechanics. These relations desribe the 
connections between quantum objects (particles) and space 
(or space-time). On a fundamental level (at Planck's scale),
there may be a complicated mixture of entanglement involving 
spatial points and quantum particles. In the large scale limit,
which governs the phenomena of elementary particles and
atoms, this mixture is effectively described by the wave function
of quantum mechanics. So, rather than representing a
strict ``either-or-not''-relation, the lines in the figures
presented above represent complex numbers, and these complex numbers
express a semi-classical limit of the entanglement
between spatial points and quantum objects. 
In this way the interference of relations can be explained,
i.e., the fact that relations can ``annihilate'' each
other by superposition. (On a more fundamental
level such an annihilation can also be explained by
opposite ``flows'' along the lines representing the
relations. One may even speculate that the relations
represent a ``flux of (quantum) information''.) 

Let $\psi_e(x)$ be the wavefunction of an electron. We
define the generalized adjacency matrix attributed to
a system consisting of spatial points and an electron
by
\begin{equation}
\label{eq2}
    A(e,x) = \psi_e(x) \, , 
\end{equation}
where $e$ refers to the ``object'' electron and $x$ 
to a labeling
of the spatial points (compare fig.\ \ref{fig2_1}).  
Thus one arrives at the following form of the generalized
``adjacency matrix'' including spatial points $x_i$ and
quantum objects $e_i$:

\begin{equation}
{\rm Ad} ~ \simeq ~ \left( \begin{array}{cc}
  A(e_i,e_j) \simeq \left( \begin{array}{c}
   \mbox{measure for} \\ \mbox{entanglement} 
  \end{array} \right) & A(e_i,x_j) \simeq \Psi_i(x_j)
  ~~ \mbox{wave function} \\
  A(e_i,x_j)^T \simeq \Psi_i(x_j)^* &
  \begin{array}{cc} A(x_i,x_j) \longrightarrow g_{\mu \nu} \\
  \mbox{(large scale limit)} \end{array} \end{array} \right)
\end{equation}

\section{The dynamics of relations}
\label{sec4}

In this section, I will speculate about the possible dynamics of 
the model introduced in the last section. In particular, I will
propose a new locality principle, based on the dynamics of
relations. At this point the notion of time enters.
In lack of a fundamental theory, I will use a
``semi-phenomenological'' ansatz and consider ``time'' 
as a sequence of discretized steps and formulate
an iterative equation for the entries of the adjacency
matrix in eq.\ \ref{eq2}. The objects themselves don't move.
Movement is interpreted as a change of relations.

For the ``propagation of relations'' it seems close at hand to 
require the following general ``locality principle'': Two quantum 
objects $e_1$ and $e_3$ can only become related, if there exists
a third quantum object $e_2$ such that $e_1$ and $e_2$ 
as well as $e_2$ and $e_3$ are already related. As this
step-weise propagation of a relation involves a single
``time-step'' (which is expected to be of the
order of the Planck time), it would appear to be ``immediate''
compared to macroscopic scales.  

As far as entanglement among quantum systems is concerned,
this locality principle can be demonstrated in a well-known
example: the measurement on an EPR-state. Consider a system
consisting of three subsystems: two entangled electrons $e_1$
and $e_2$ in an EPR-state, and a measurement apparatus (assumed
to be in the state $|0\rangle_3$ initially). Hence, the initial
state is:
\begin{equation}
\label{EPR}
  |\Psi\rangle_{\rm i} = \frac{1}{\sqrt{2}} \Big(
     | \uparrow \rangle_1 | \downarrow \rangle_2 -
     | \downarrow \rangle_1 |\uparrow \rangle_2 \Big)
     |0\rangle_3 ~. 
\end{equation}
The apparatus now
performs a measurement on electron 2 (by an ordinary local
interaction involving a flow of energy). This leads to
entanglement between the apparatus and electron 2.
If electron 2 were not entangled but in the pure state 
$\frac{1}{\sqrt{2}}(|\uparrow\rangle_2 - |\downarrow\rangle_2)$, 
this process could be described by:
\begin{equation}
\frac{1}{\sqrt{2}} \Big( |\uparrow \rangle_2 -
  |\downarrow \rangle_2\Big) |0\rangle_3 \longrightarrow
  \frac{1}{\sqrt{2}} \Big( |\uparrow\rangle_2
  |+\rangle_3 - |\downarrow\rangle_2 |-\rangle_3\Big)~.
\end{equation}
As an immediate consequence of this interaction, the already 
existing entanglement between electron 1 and electron 2
``swapps over'' to the measurement apparatus and the
final state reads:
\begin{equation}
\label{final}
 |\Psi\rangle_{\rm f} = \frac{1}{\sqrt{2}} \Big(
 |\uparrow\rangle_1 |\downarrow\rangle_2 |-\rangle_3
 - |\downarrow\rangle_1 |\uparrow\rangle_2
   |+\rangle_3 \Big)~,
\end{equation}
expressing entanglement of all three subsystems.

As a side-remark I should like to notice that the
relational picture in its present form does not solve
the measurement problem, i.e., it does not describe or
explain the ``collaps'' of the state \ref{final}:
\begin{equation}
\label{collaps}
 |\Psi\rangle_{\rm f} = \frac{1}{\sqrt{2}} \Big(
 |\uparrow\rangle_1 |\downarrow\rangle_2 |-\rangle_3
 - |\downarrow\rangle_1 |\uparrow\rangle_2
   |+\rangle_3 \Big) \longrightarrow
  \left\{ \begin{array}{c}
  |\uparrow\rangle_1 |\downarrow\rangle_2 |-\rangle_3 \\[0.2cm]
   \mbox{or} \\[0.2cm]
  |\downarrow\rangle_1 |\uparrow\rangle_2 |+\rangle_3 
   \end{array} \right. 
\end{equation}
It is this collaps, not the interaction between electron 2
and the measurement apparatus, which destroys the
entanglement and leads to a (non-entangled) product state.

As mentioned before, there is a slight difference between 
the relational picture
of the propagation of relations and the standard quantum
mechanical description: In the standard interpretation, the
entanglement between the measurement apparatus and electron
1 evolves parallel to the entanglement between the apparatus
and electron 2. In the relational picture presented so far,
the entanglement between the apparatus and electron 1 occurs
``one time-step after'' the entanglement between the
apparatus and electron 2 has been established. Apart from
the fact that the the build-up of entanglement is not an
instantaneous ``jump'' but rather a continuous process
initiated by the interaction, this difference of ``one time-step''
(which is assumed to be of the order of Planck's time,
i.e., $10^{-44}$\,s) cannot be directly observable. 
However, we do observe this difference indirectly as 
the limit on the propagation velocity (the speed of light) of
objects. It would be interesting and at the same time a 
first test of this relational picture, if this finite propatation
velocity for entanglement could be observed more directly,
e.g.\ in condensed matter systems or other many particle 
systems.  
 
In order to be more explicit about the dynamics of the other
relations (the wave function and the spatial part of the 
generalized adjacency matrix), we need the adjacency
matrix for the spatial points, i.e., we need to
know $A(x,y)$. A simplifying assumption treats the spatial
relations as fixed, i.e., space points do not take part
in the evolution of the system and the matrix $A(x,y)$ does
not change. In a more fundamental model, this will no
longer be true: the relations among spatial points
will fluctuate and change; however, the large scale limit 
-- the metric -- will remain constant. This puts severe
restrictions on the fundamental dynamics of relations
among spatial points.

Let $A(x,y)$ be the adjacency matrix of the graph
(eq.\ \ref{eq1}), then we can define the so-called 
graph-Laplacian:
\begin{equation}
  \Delta(x,y) = A(x,y) - V(x,y) \, , 
\end{equation}
where 
\begin{equation}
   V(x,y) = \left( \sum_z A(x,z) \right)\, \delta(x,y)  
\end{equation}
is the diagonal valence matrix of the graph ($\delta(x,y)$
beeing the discrete Kronecker-Delta). $\Delta(x,y)$
is the discretized analogue of the standard Laplacian on
a manifold. The discretized (free) Schr\"odinger equation
for the relations $A(e,x;t)$ of particle $e$ now reads:
\begin{equation}
  {\rm i} A(e,x;t+1) = {\rm i} A(e,x;t) - 
        \mu \sum_y \Delta(x,y)\, A(e,y;t) \, , 
\end{equation}
where $\mu$ is some constant which in the
continuum limit has to be renormalized
to the inverse of the mass $m$ of the particle. 
This may be considered as a discretized version of 
quantum mechanics, and in the continuum limit the
wavefunction obeys Schr\"odinger's equation.

Employing this formalism, we also arrive at an interesting
interpretation of Feynman's ``summation over paths''.
In quantum mechanics, the general solution of 
Schr\"ordinger's equation can be written in the form:
\begin{equation}
  \psi(x,t) = \int \! {\rm d}y \, K(x,y;t)\, \psi(y,0)~.
\end{equation}
Formally, the Kernel $K(x,y;t)$ can be represented
as a sum over all paths of ``length'' $t$ from point 
$y$ to point $x$, where each path is ``weighted'' by a
phase depending on its classical action. 
Feynman's ``summation over paths'', or,
more generally, ``summation over histories'' is
an intuitive expression.
The formalism requires, however, that ``a particle 
propagates along path 1 AND path 2 AND path 3 ...'',
which is difficult to understand.
If we interprete the wave function 
as encoding the relations between a particle and space,
Feynman's integral reads ``propagation of relation 1
along path 1 AND relation 2 along path 2 AND ...''.
(According to the locality principle, relations can 
also split or merge; this finally leads to the sum
over ``all'' paths.) Interpreted in this way, the
sum over histories appears much less ``mysterious''.

\section{Are all ``properties'' relational?}
\label{sec5}

In standard quantum mechanics we are free to
choose a basis in Hilbert space. If we denote by $|\psi\rangle$
some vector in this Hilbert space representing the state of
the system, Schr\"odinger's wave function 
is usually expressed in the position
base: $\psi(x)=\langle x|\psi\rangle$. However, 
we can also choose the momentum base instead 
($\tilde{\psi}(p)=\langle p|\psi\rangle$) or the
(orthonormal) base defined by the eigenvectors of
any other observable. 

In the relational picture outline in the previous
sections, the position basis is distinguished. 
We have assumed that the ``spatial points''
have an objective reality in the same sense as the
particles have an objective reality. The relation
between these two objects is expressed by 
$\psi_e(x) =\langle x|\psi\rangle$. The
transformation to the momentum representation
is obtained by a Fourier transformation like 
in standard quantum mechanics. However, it is not
clear in what sense $\langle p |\psi\rangle$ 
can be interpreted as
a relation between two ``objects''. In this picture
there is no objective ``momentum point'', although
formally one can represent the relational picture
in any other basis.
In an even more general setting one would like to
gain back the lost symmetry and interprete the
transition amplitudes $\langle a|\psi\rangle$ as
a relation between an ``object'' $\psi$ and an 
``object'' $a$.

In its present formulation, the model presented here
is far from this goal. However,
I will show that even in standard
quantum mechanics and quantum field theory, many
``properties'' of particles emerge through 
relations between this particle and its environment.
The purpose of this section is to show that
relational properties are nothing new even within 
standard theories in physics.

One of these properties is the mass $m$ of a particle
in the standard model of elementary particles. 
Usually one starts from a Lagrangian in which all
particles are treated
as massless (see, e.g., \cite{Weinberg}). As a result
of spontaneous symmetry breaking, the Higgs-field
aquires a non-vanishing expectation value, and
due to the Yukawa-type interaction between the 
other particles in the model and this background
Higgs field the particles aquire
a mass.

Yet, we do not have to employ the mechanism
of symmetry breaking in order to find relational
properties in quantum field theory. The observed
mass $m$ and charge $e$ of the electron in 
quantum electrodynamics are not properties of
the ``bare'' electron. In the standard interpretation,
these properties are determined by the interactions
between the electron and the quantum fluctuations
of the environment (expressed essentially by the 
higher order corrections of the two- and
four-point functions in Feynman's perturbation
theory). In standard QED this leads to the
renormalization of the mass and the charge of the
electron.

As a third property, I should like to mention
the formalism by which ``spin'' is usually described 
in quantum mechanics. The Hilbert space of 
square-integrable functions handles the spatial 
degrees of freedom of a particle, and in addition
a two-dimensional complex vector space 
refers to the spin of the particle. This tensor space
construction expresses the fact that the spin 
degrees of freedom of a particle are independent of
the spatial degrees of freedom. The spin state and
the spatial state of the particle can even be entangled: 
In the neutron interference experiments \cite{Hasegawa}, 
the intermediate state of the neutron can be written as:
\begin{equation}
  |\psi\rangle = \frac{1}{\sqrt{2}} \Big(
    \psi_1(x)|\uparrow \rangle +
    \psi_2(x)|\downarrow\rangle \Big)\, ,
\end{equation}
where $\psi_1$ and $\psi_2$ refer to two
different paths of the particle through the
interferometer. A similar situation occurs
when a photon hits a polarization beam splitter
which reflects photons of one polarization and
transmits photons of the orthogonal polarization.
Thus, the formal description of spatial and spinorial
degrees of freedom of a particle resemble the
description of multi-particle systems. 

\section{Relational models of space-time}
\label{sec7}

In this final section, I should like to make some
remarks about the generalization of
the above described model of relations between particles
and spatial points towards a theory of relations between
events and space-time points. At present, these
remarks are very speculative. 

In 1990, Rudolf Haag raised a very fundamental question 
\cite{Haag1}: What are elementary events? Despite the
fundamental importance the concept of events has for
general relativity, it has never been formalized within
the context of quantum theory or quantum field theory.
(A more recent update of his ideas can be found in \cite{Haag2}.)
We know that within the framework of perturbation theory, we can 
express the transition amplitudes of quantum field theory 
by a sum over Feynman integrals and, in a graphical notation,
by a sum over Feynman graphs. These Feynman graphs show
elementary events (like the emission or absorption of a
photon by an electron in QED), but these events are not
``factual'', they have to be considered as
``virtual'' events or ``possibilities''. Haag studies the
transition from possibilities to factuality by looking at
increasing clusters of virtual events, finally arriving
at partitions of our universe into factual subsystems.

In the section \ref{sec4}, I have already explained that
the ``summation over histories'' (including a summation over
all possible kinds and locations of elementary events allowed
by the classical action) appears much more natural in a
relational interpretation. The ``elementary events'' do not
involve actual particles propagating through space, but only 
certain relations between
particles and spatial points. In this way there is no need
to explain a transition from possibilities to facts, but
all relations may be considered as factual. Strictly speaking, 
however, the ``event'' of, say, the scattering of two electrons 
is distributed all over space-time.

On a closer inspection, however, the generalization
of the scenario described in the previous sections from
``spatial points'' to ``space-time events'' turns out
to be more problematic. First, in order to reproduce the standard
results of quantum field theory, a certain event can be related
to all other space-time-events, independent of whether the two
events are space-like or time-like (i.e., independent of whether
they are causally related or not). This does not violate the
principle of causality as long as the relational weights reproduce
the known causal Green functions (whose real parts are also 
non-zero for space-like events). In a large scale limit
(where ``large'' can mean anything larger than Planck's
scale) these relations can still reproduce other 
discretized models of space-time (like causal sets \cite{Sorkin}
or random networks \cite{Requardt}).

While the relational picture can reproduce the contribution
of a single Feynman graph (including the integrations over
all internal space-time points of elementary events), the
complete theory requires also a sum over all possible Feynman
graphs, i.e.\ over all possible combinations of elementary
events. In the present picture, this summation cannot be 
replaced by a single set of relations. The ``superposition
principle for a second-quantized theory'' turns out to be
more subtle and may require an even more general formalism
of relations.  

\section{Summary}

It is argued that if we treat the position of a particle
not as an embedding into some background space but as an 
expression of the relations between this particle 
and spatial points, and if we interprete
the wave function in quantum mechanics as encoding these
relations, we arrive at a relational interpretation of
quantum mechanics which not only solves some of its 
``mysteries'' (the particle-wave duality or the spooky
action at a distance) but which might also be a way to
circumvent the restrictions set by Bell's inequalities 
on ``local realism'' in a hidden
variable theory. The aim of the present article was to
show that in such a framework the formalism of quantum 
mechanics remains almost unchanged, but the interpretation
of many expressions becomes more natural.

\end{document}